\documentstyle[emulateapj,psfig]{article}

\begin{document}

\title{The Quiescent Accretion Disk in IP Peg at Near-Infrared Wavelengths}

\author{C. S. Froning, E. L. Robinson, William F. Welsh}
\authoremail{cyndi@imp.as.utexas.edu, elr@pisces.as.utexas.edu,
wfw@aries.as.utexas.edu} 
\affil{Department of Astronomy, University of Texas, Austin 78712}
\author{Janet H. Wood}
\authoremail{jhw@astro.keele.ac.uk}
\affil{Department of Physics, Keele University, Keele, Staffordshire
ST5 5BG}

\begin{abstract}
We present near-infrared, H-band (1.45 - 1.85 $\mu$m) observations of
the eclipsing dwarf nova, IP Peg, in quiescence.  The light curves are
composed of ellipsoidal variations from the late-type secondary star
and emission from the accretion disk and the bright spot.  The light
curves have two eclipses: a primary eclipse of the accretion disk and
the bright spot by the companion star, and a secondary eclipse of the
companion star by the disk.  The ellipsoidal variations of the
secondary star were modeled and subtracted from the data.  The
resulting light curve shows a pronounced double-hump variation.  The
double-hump profile resembles those seen in the light curves of WZ Sge
and AL Com and likely originates in the accretion disk.  The primary
eclipse was modeled using maximum entropy disk mapping techniques.
The accretion disk has a flat intensity distribution and a cool
brightness temperature ($T_{br} \simeq $ 3000 K) in the near-infrared.
Superimposed on the face of the disk is the bright spot ($T_{br}
\simeq $ 10,000 K); the position of the bright spot is different from
the observed range of visible bright spot positions.  The
near-infrared accretion disk flux is dominated by optically thin
emission.  The secondary eclipse indicates the presence of some
occulting medium in the disk, but the eclipse depth is too shallow to
be caused by a fully opaque accretion disk.

\end{abstract}


\section{Introduction}
%
Observations at near-infrared wavelengths (NIR) are well-suited for
studying the companion star and the outer accretion disk in
cataclysmic variables (CVs) and other compact binaries.  The spectral
type of the late-type secondary star can be determined from its NIR
colors or its absorption line strengths, although both methods usually
require an estimate of the contribution of the accretion disk to the
near-infrared flux (\cite{berriman1985,dhillon1995}).  In
high-inclination compact binary systems, the NIR light curves
typically show ellipsoidal variations from the Roche-lobe shaped
secondary star.  The ellipsoidal variations constrain the inclination
and mass ratio of the binary and, when combined with the observed mass
function, place limits on the mass of the primary object, an important
means of modeling black hole binary systems (e.g. Haswell et
al.\ 1993, Shahbaz et al.\ 1994) \markcite{haswell1993,shahbaz1994a}.

\begin{table*}
\begin{center}
\begin{tabular}{lcc}					
\multicolumn{3}{c}{Table 1:  Observations of IP Peg} \\ \tableline\tableline
Date (UT) 	&	Exposures &	Integration Time \\ \tableline
							\\
1993 Sept 4 	&	260  &		10s		\\
1993 Sept 5 	&	381  &		10s		\\
1993 Sept 7 	&	646  &		10s		\\
1994 Sept 28 	&       1404 &		5s		\\
1994 Sept 29 	&       2346 &		5s		\\
1994 Sept 30 	&       2548 &		5s		\\
1994 Oct 25 	&       1795 &		5s		\\
1994 Oct 27 	&       155  &		5s		\\ \tableline
\end{tabular}
\end{center}
\label{logobs}
\end{table*}
Near-infrared photometry of CVs also probes their accretion disks.
The visible and ultraviolet light curves of CVs are dominated by flux
from the accretion disk, the white dwarf, and the bright spot where
the mass transfer stream impacts the disk.  The disk flux seen at
these wavelengths emanates predominantly from hotter, inner annuli,
the disk/white dwarf boundary layer, and the disk chromosphere.
Near-infrared data supplement these shorter wavelength observations by
probing the cool disk at larger radii.  It is in precisely these
regions of the disk, where material accreted from the companion star
accumulates in quiescence, that many physical phenomena of interest
take place, including the initiation of some dwarf novae outbursts.
High-inclination CVs show eclipses of the accretion disk at inferior
conjunction of the secondary star; in the near-infrared, secondary
eclipses of the companion star by the disk are also seen
(\cite{bailey1981b,sherrington1982}).  Flux-ratio diagrams of
near-infrared emission from CVs indicate the presence of both opaque
and transparent material in the accretion disk, and show that the
fraction of optically thick to optically thin gas varies from system
to system (\cite{berriman1985}).  A combination of multicolor NIR
light curves and spectra can be used to determine the relative
contributions of the secondary star and the accretion disk to the
total flux, and the amount of optically thick and optically thin
emission from the disk.

In nearly edge-on systems, where the companion star occults the
primary star and the accretion disk, the shape of the eclipse contains
information on the pattern of emissivity across the disk.  Maximum
entropy modeling methods have been successfully used at optical and UV
wavelengths to construct maps of the accretion disk intensity and
determine its radial brightness temperature profile (see Horne
1993\markcite{horne1993} for a review).  In dwarf novae in outburst,
visible disk maps show brightness temperature (\(T_{br}\)) profiles
roughly consistent with optically thick, steady-state emission
(e.g. Horne and Cook 1985; Rutten et
al.\ 1992\markcite{horne1985b,rutten1992a}).  Maps of quiescent dwarf
novae, however, show flat \(T_{br}\) profiles and accretion disks that
are far from steady-state (\cite{wood1986a,wood1992}). In order to
extend multiwavelength modeling of accretion disks, we present the
first near-infrared map of a disk, in the cataclysmic variable,
IP~Peg.

IP Peg is a dwarf nova consisting of a mass-donating, M-dwarf
secondary star and an accretion disk around a white dwarf primary
star.  It undergoes regular outbursts every few months in which it
brightens by approximately two magnitudes in the visible. IP Peg has
an orbital period of 3.8 hours.  With its high inclination, it is one
of the few dwarf novae above the period gap with eclipses of the white
dwarf, accretion disk and bright spot.  Visible eclipse timings have
constrained the geometry of the system, although the fits are
complicated by blending of the eclipse features due to the bright spot
and white dwarf ingress (\cite{wood1986b,wolf1993}).  The large
contribution of the bright spot to the eclipse profile has also
precluded any quiescent, visible maps of the accretion disk.  Szkody
\& Mateo (1986)\markcite{szkody1986} observed IP Peg in the
near-infrared.  They obtained mean J, H and K colors and their light
curves show ellipsoidal variations and both primary and secondary
eclipses.

In this paper, we present H-band light curves of IP Peg, fits to the
ellipsoidal variations of the secondary star, and a map of the
accretion disk.  Section 2 summarizes the observations and data
reduction.  Section 3.1 discusses the morphology of the light curves,
while Section 3.2 describes modelling the ellipsoidal variations.  We
describe the eclipse mapping method and present the quiescent disk map
in Section 3.3, and consider the dependence of the results on the
choice of modeling parameters in Section 3.4.  Section 4 discusses the
results, and concluding remarks are presented in Section 5.

\section{Observations and Data Reduction}
%
%
We observed IP Peg for three nights in 1993 September and five nights
in 1994 September and October using the infrared imaging camera,
ROKCAM (\cite{colome1993}), on the 2.7-m telescope at McDonald
Observatory.  We obtained 15 hours of H-band data; the observations
are summarized in Table 1. IP Peg was in quiescence during the
observations; the next outbursts began on 1993 October 25 and on 1994
December 7 (\cite{aavso277,aavso291}).  We observed IP Peg and a
nearby field star, located approximately 34 arcseconds to the
southwest, and measured the sky background by nodding the telescope in
a grid of nine positions.

The data were reduced using the standard IRAF data reduction packages
and the DAOPHOT aperture photometry routines.  The data initially
showed evidence of an instrumental artifact, causing variations of up
to $\pm0.1$ mag in both the target and the comparison stars.  The
variations were correlated with the grid position of the star on the
array, which we attributed to a poor match between the dome flats and
the actual sensitivity variations across the ROKCAM chip.  We removed
most of the instrumental signature by applying corrective terms to the
calculated instrumental magnitudes. The correction for each grid
position and each star was computed from the difference between the
mean magnitude of the star and its mean magnitude in a given grid
position for each night.  We then subtracted the field star magnitude
from the IP Peg magnitude in each frame to correct for atmospheric
effects.

The shapes of the resulting light curves did not vary appreciably from
one night to the next, so we combined the data into three mean light
curves: 1993 September, 1994 September, and 1994 October.  The
individual nights were shifted to have the same mean magnitude before
combining (a typical shift of 0.008 mag).  The data from 1994 October
27 were not used in the mean 1994 October light curve because IP Peg
appeared to be fainter on that night, and with the limited data, it was
impossible to determine whether this was real or an instrumental
effect.  We converted the data from magnitude to arbitrary intensity
units and binned them using phase bins equal to 0.5\% of the orbital
period.  We used the Wolf et al.\ (1993)\markcite{wolf1993} ephemeris,
in which phase zero corresponds to the phase of white dwarf egress.  A
bin size of 0.005 is equal to a time interval of 1.14 minutes.

We calibrated the data using standard star observations from 1994
September 30 (\cite{elias1982}).  Since only two standard stars were
observed, we fit the data using a simple transformation equation with
a fixed extinction coefficient and no color terms (\cite{allen1976}).
The mean H magnitude for IP Peg on 1994 September 30 was
$12.14\pm0.11$, a value consistent with previous observations
(\cite{szkody1986}).  The conversion from magnitude to intensity units
was calibrated using a flux calibration of $10^{6}$ mJy for a
zero-magnitude M-dwarf star (\cite{berriman1987}).  The standard
deviation of the mean after flux calibration averaged $\pm0.12$,
$\pm0.06$, and $\pm0.11$ mJy for 1993 September, 1994 September and
1994 October, respectively.  The uncertainties were dominated by
systematic effects such as the sensitivity variations of pixels across
the array, so we set the error bars in each light curve equal to the
mean error.  We tripled the error bar on a single bin near primary
minimum in the 1994 September light curve ($\phi$ = 0.0625; see
Figure~\ref{modfit}), because the deviation of this point from the
surrounding points is not real, but a remnant of the instrumental
calibration error.

\begin{figure*}
	\hspace*{.85in}\psfig{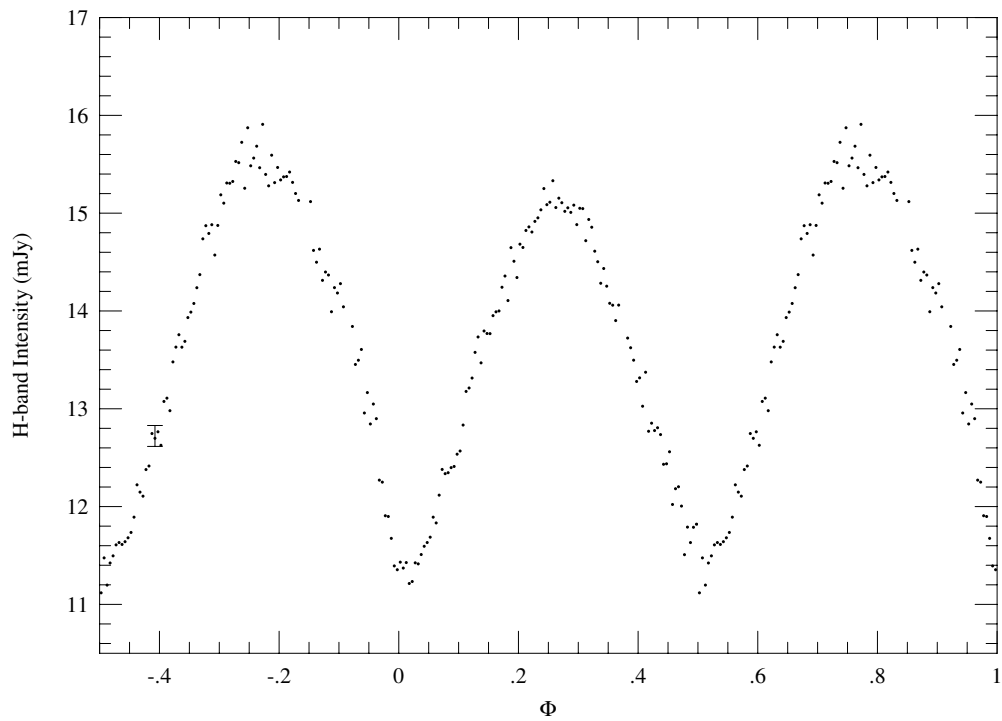}
 	\caption{1993 September mean light curve.  The data points are
binned to 0.005 in orbital phase.  The error bar shown represents the
mean systematic errors.  The flux calibration uncertainty is $\pm0.6$
mJy.}  \label{fsept93}
\end{figure*}

\begin{figure*}
	\hspace*{0.85in}\psfig{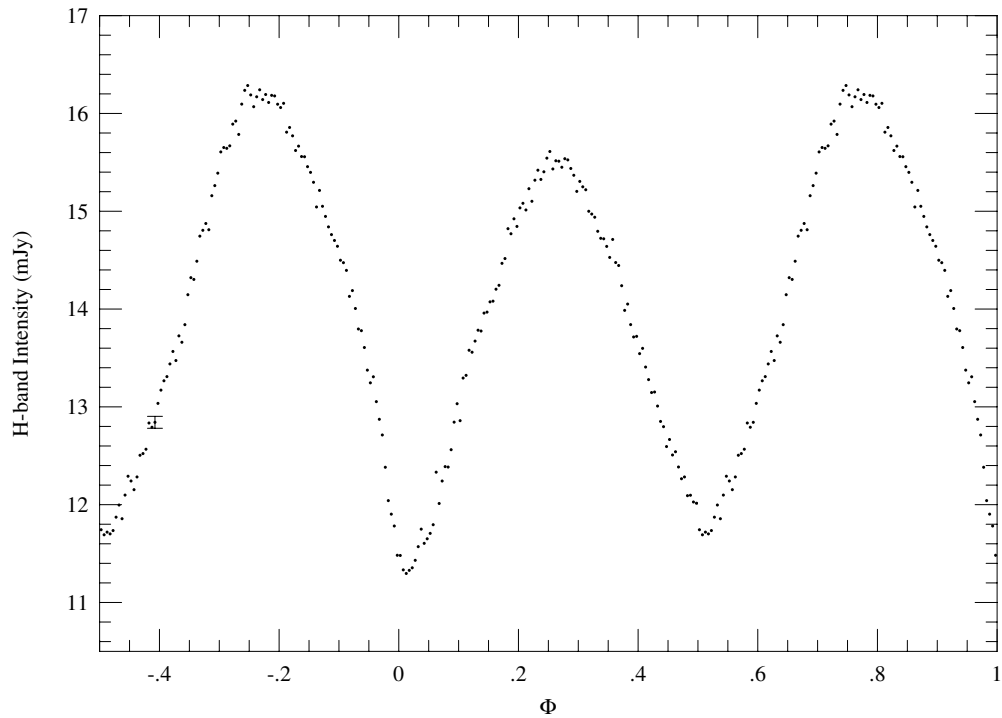}
  	\caption{1994 September mean light curve.  The data points are
  	binned to 0.005 in orbital phase.  The error bar shown
  	represents the mean systematic errors. The flux calibration
  	uncertainty is $\pm0.6$ mJy.}
 	\label{fsept94}
\end{figure*}

\begin{figure*}
	\hspace*{0.85in}\psfig{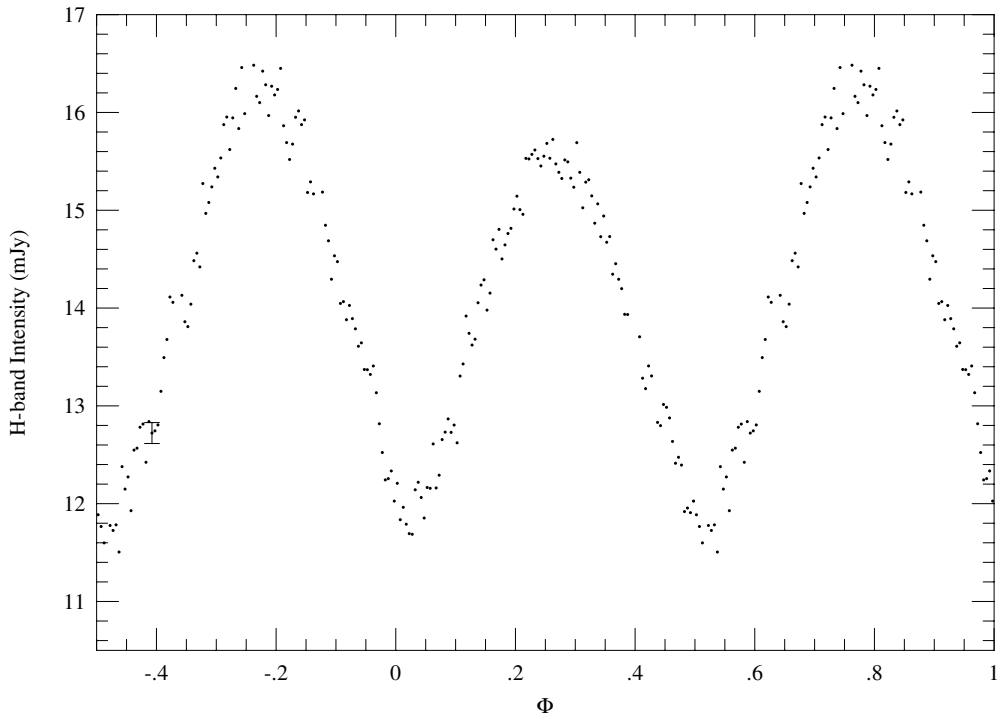}
 	\caption{1994 October mean light curve.  The data points are
binned to 0.005 in orbital phase.  The error bar shown represents the
mean systematic errors.  The flux calibration uncertainty is
$\pm0.6$ mJy.}
	\label{foct94}
\end{figure*}
\section{Analysis of the Quiescent Light Curves}                   
%
%
\subsection{Morphology}

The mean H-band light curves of IP Peg are shown in
Figures~\ref{fsept93} through~\ref{foct94}.  Due to the difficulty in
measuring the white dwarf ingress in IP Peg, the published ephemeris
designates $\phi$ = 0 as the phase of white dwarf egress, but
convention -- and the maximum entropy modeling program -- place the
point of zero phase at inferior conjunction of the secondary star.
The observed light curves can be shifted to the conventional zero
phase using the known width of the white dwarf eclipse
(\cite{wood1986b}).  Due to changes in IP~Peg's orbital period,
however, the linear ephemeris is imprecise.  As a result, we shifted
the light curves by half the white dwarf eclipse width plus an
additional amount ($\Delta \phi$ = 0.027) to correct for deviations
from the ephemeris.  This additional shift was determined while
mapping the accretion disk (see Section 3.3) and applied iteratively
to the data.  The orbital phasing shown in Figures~\ref{fsept93}
through~\ref{foct94} incorporates this correction.

All three light curves are dominated by what appears to be ellipsoidal
variations from the late-type companion star, but it is clear that the
secondary star is not the only near-infrared emitter.  The peak in the
light curve near $\phi$ = 0.75 is greater than the peak near $\phi$ =
0.25; typically, this phenomenon is indicative of beamed emission from
the bright spot.  The primary eclipse of the accretion disk is visible
in the 1993 September and 1994 September light curves.  A primary
eclipse is probably present in the 1994 October data, but it is less
obvious in that light curve.  The 1994 September light curve also
shows bright spot ingress and egress features (near $\phi$ = 0.9 and
0.1 in Figure~\ref{fsept94} and more clearly in Figure~\ref{dhump}).
The primary eclipse is not deep, suggesting that the integrated flux
from the accretion disk in IP Peg is small in the H-band.  The second
minimum in the data near $\phi$ = 0.5 is caused partially by the
ellipsoidal variations, but a secondary eclipse of the companion star
by the accretion disk may also be present.

The 1993 September and 1994 September light curves are morphologically
similar.  The flux at primary minimum is nearly the same, but the 1993
data has smaller peak-to-peak variations, indicating that the
accretion disk and/or the bright spot fluctuate in brightness.  The
1994 October light curve is significantly different from the September
data: the primary eclipse is shallower and the eclipse minimum is
shifted to a later orbital phase.  The secular variations in IP Peg
precluded combining all the data into a single light curve.  Since the
1994 September light curve has the best signal-to-noise ratio, we
concentrated on modeling it.  All further references to the data,
unless stated otherwise, refer to Figure~\ref{fsept94}.

\subsection{Modeling the Ellipsoidal Variations}

The first step in analyzing the light curve required modeling and
removing the contribution of the secondary star.  The ellipsoidal
variations were modeled using a light-curve synthesis program
(\cite{zhang1986,haswell1993}). The program calculates the flux from a
Roche lobe-filling secondary star by dividing the surface of the star
into a grid.  The flux at each grid position is modified to account
for the effects of gravity-darkening and limb-darkening:

\begin{equation}
T(r,\theta,\phi) =
T_{pole}\left[\frac{g(r,\theta,\phi)}{g_{pole}}\right]^{\beta}
\label{eqnt}
\end{equation}

and

\begin{equation}
I = I_{0}(1-u+u\cos\gamma)	\label{eqni}
\end{equation}

\noindent
where $\beta$ and $u$ are the gravity darkening and linear
limb-darkening coefficients, respectively, $T_{pole}$ and $g_{pole}$
are the temperature and surface gravity at the pole of the star,
$I_{0}$ is the intensity emitted normal to the surface of the star,
and $\gamma$ is the angle between the line of sight and the normal to
the surface.  

The output of the program is a light curve of flux versus orbital
phase in arbitrary flux units.  Traditionally, the program has been
used to create model light curves by varying the free parameters until
the fit to the observed data has been optimized.  The free parameters
for a model of the secondary star are the mass ratio, $q$, the orbital
inclination, $\imath$, $T_{pole}$, $\beta$, and $u$.  Since the
secondary star is not the only source of modulation in the light curve
of IP Peg, we initially fit the model light curve to the data in the
orbital phases between $\phi$ = 0.1 and 0.4.  We assumed that the flux
at these phases consists of the ellipsoidal variations plus a constant
disk component, that is:

\begin{equation}
F_{obs}(\phi) = c \cdot F_{2}(\phi) + F_{d}  \label{eqnf}
\end{equation}

\noindent
where $F_{obs}$ is the observed data, $F_{2}$ is the flux from the
secondary star and $F_{d}$ is the contribution of the accretion disk.
The constant term, $c$, scales the modeled secondary star flux to the
data.  We then solved for $c$ and $F_{d}$ using a least-squares fit to
determine the fractional contributions of the secondary star and the
disk to the observed data.  This method assumed that the uneclipsed
disk flux is constant over the observing period and that --- consistent
with the visible light curves of IP Peg --- the bright spot does not
contribute to the data at these orbital phases (\cite{wood1986b}).

To create the model light curve of the secondary star, $F_{2}(\phi)$,
we initially set $q$ = 0.49 and $\imath$ = 80$\fdg$9, values derived
from the visible eclipses of IP Peg (\cite{wood1986b}).  The spectral
type of the secondary star and its temperature were determined from
the observed (H-K) color and set to $T_{pole}$ = 3050 K
(\cite{szkody1986,leggett1992}).  The gravity darkening coefficient
was set to $\beta$ = 0.05 (\cite{sarna1989}), and the limb-darkening
coefficient was extrapolated from Wade \& Rucinski (1985)
\markcite{wade1985} and set to $u$ = 0.30.  We also varied the
parameters to cover the range of physically permitted models for the
secondary star, varying $q$ from 0.35 to 0.6, $\imath$ from 89$\fdg$5
to 79$^{\circ}$, $T_{pole}$ from 2800 to 3200 K, $u$ from 0.2 to 0.5,
and $\beta$ from 0.05 to 0.08.  We were unable to fit a viable model
to the observed data for any of the parameter values.  Every fit to
$\phi$ = 0.1 to 0.4 generated more modeled flux than observed flux at
other orbital phases.  This was particularly evident at mid-eclipse,
$\phi$ = 0, where the modeled secondary star flux must be equal to or
less than the observed flux, and instead was higher.  The best fits to
Equation~\ref{eqnf} for the secondary star models also generated
negative, unphysical values for the disk flux, $F_{d}$.  The failure
of the ellipsoidal variations plus a constant disk component to fit
the data for $\phi$ = 0.1 -- 0.4 demonstrates that an additional,
phase-dependent source contributes to the IP Peg light curve in the
near-infrared, partly mimicking an ellipsoidal variation.

\begin{figure*}
	\hspace*{.85in}\psfig{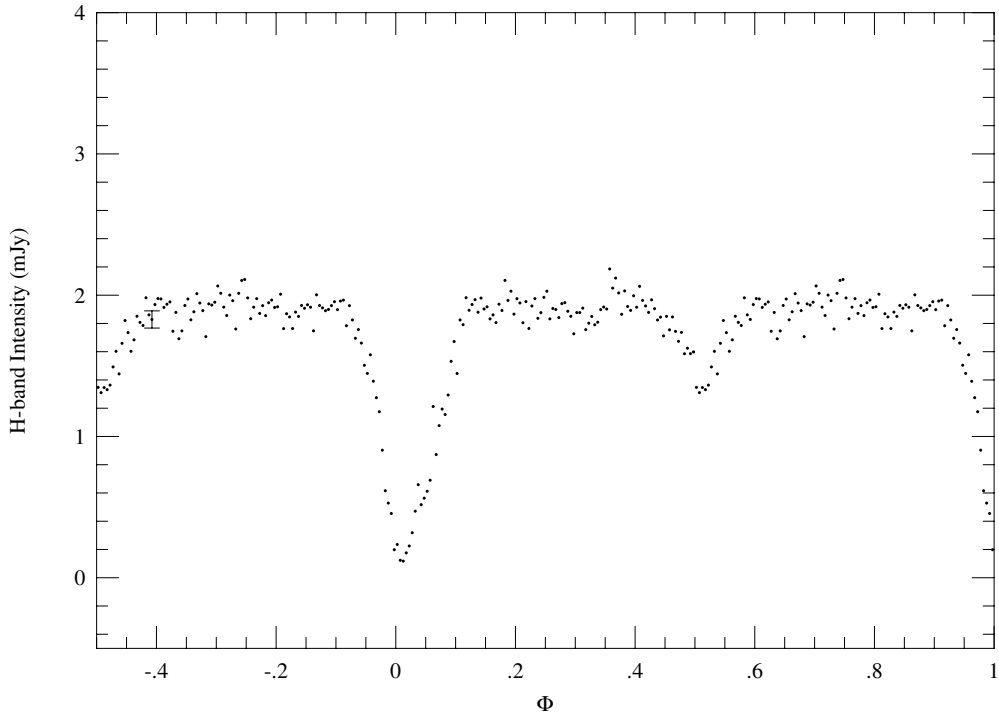}
	\caption{The 1994 September light curve after rectification by
	the double-sine curve.  The primary eclipse was subsequently
	modeled using the maximum entropy method (MEM) of accretion
	disk mapping.}  \label{sub}
\end{figure*}

Since the near-infrared light curve alone is insufficient to constrain
the contribution of the secondary star, we abandoned the use of
Equation~\ref{eqnf} and calculated the model for the ellipsoidal
variations using best-guess parameter values obtained from previous
visible and near-infrared observations: $q$ = 0.49, $\imath$ =
80$\fdg$9, and $T_{pole}$ = 3050 K.  The model light curve was then
scaled to the data at the primary eclipse.  The scaling was done
iteratively.  First, we assumed that the secondary star is the only
source of flux at $\phi$ = 0 and set the model flux equal to the
observed flux at this phase.  After modeling the accretion disk, we
determined that 8\% of the area of the accretion disk remains
unocculted at primary minimum (i.e. the sides of the disk stick out at
mid-eclipse).  Using the primary eclipse depth (see Figure~\ref{sub})
and assuming a roughly uniform distribution of flux in the disk, we
determined that the unocculted disk emits 0.15 mJy at minimum light.
Accordingly, we rescaled the model to the observed data assuming 0.15
mJy of unocculted light at primary minimum.

For the final model, we improved the light curve synthesis program via
the addition of improved limb-darkening coefficients and specific
intensities for the secondary star ($I_{0}$).  We obtained these
parameters using the Allard model atmospheres for cool M dwarf stars
(\cite{allard1995}).  For the secondary star in IP Peg, we assumed T =
3000 K, log g = 4.5, and solar metallicity.  The resulting linear
limb-darkening coefficient in the H-band is $u\simeq 0.20$.  The
limb-darkening profile in the H-band is decidedly non-linear, however,
so the limb-darkening equation (Equation~\ref{eqni}) was modified to
use a quadratic approximation:

\begin{equation}
I = I_{0} (1 - a (1-\cos\gamma) -b(1-\cos\gamma)^{2})  \label{eqni2}
\end{equation}

\noindent
where $a$ and $b$ are the quadratic limb-darkening coefficients.  The
coefficients were derived from the model atmospheres following the
method outlined in Wade \& Rucinski (1985) \markcite{wade1985}.  For
IP Peg, the coefficients were $a$ = 0.022 and $b$ = 0.321.  Finally,
we tested the effects of including irradiation of the secondary star
in the model (assuming a 15,000 K white dwarf; Marsh
1988\markcite{marsh1988}).  The change in the depth of the ellipsoidal
dip at $\phi$ = 0.5 was 0.04 mJy, or 0.36\%.  We neglected the small
effect of irradiation to avoid additional model-dependent choices for
the temperature of the white dwarf and the albedo of the secondary
star.

\begin{figure*}
	\hspace*{.85in}\psfig{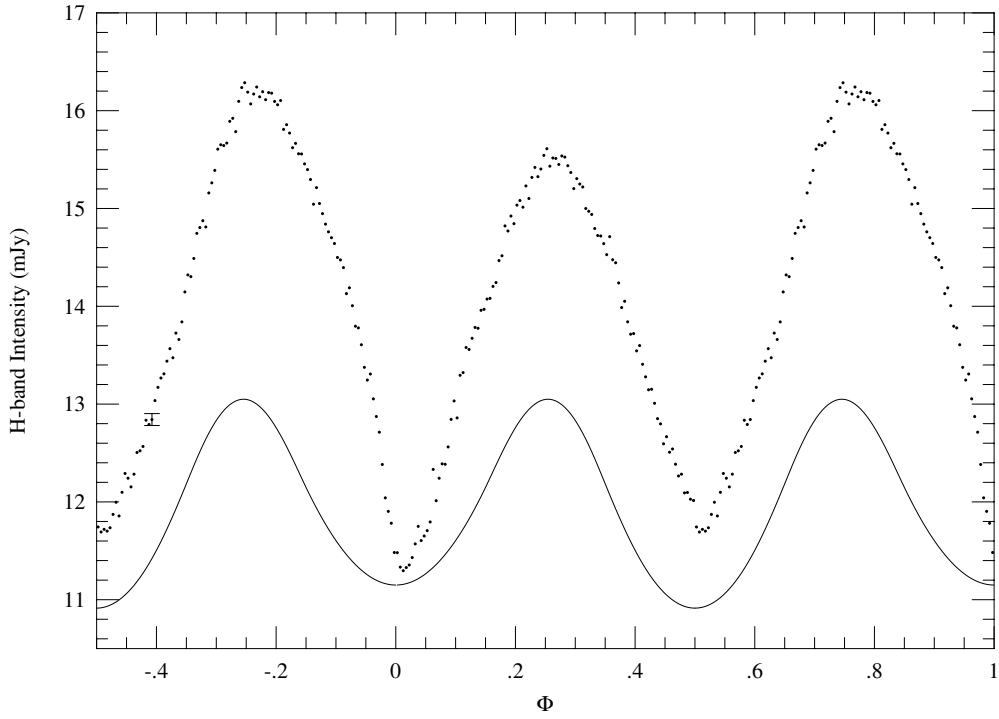}
	\caption{The 1994 September light curve repeated with the
	modeled contribution of the secondary star superposed.  The
	modeled ellipsoidal variations are scaled to the observed data
	assuming a small amount (0.15 mJy) of unocculted disk flux at
	primary eclipse minimum.}  
	\label{fit}
\end{figure*}
\begin{figure*}
	\hspace*{.85in}\psfig{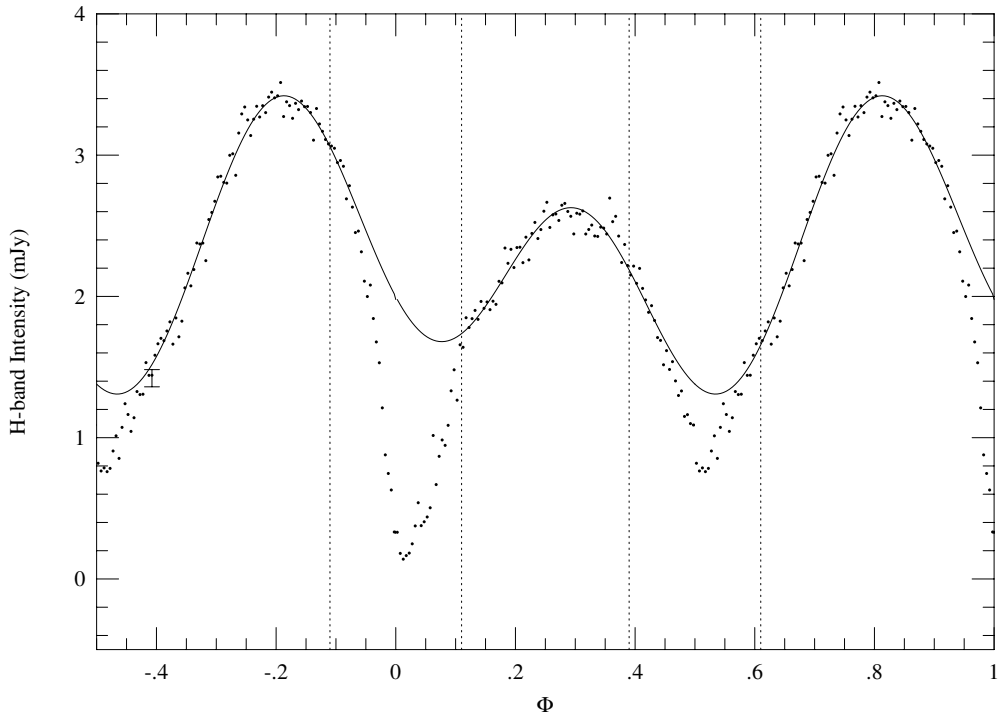}
 	\caption{The 1994 September light curve after the modeled flux
 	contribution of the secondary star has been subtracted.  The
 	solid line is the double-sine fit to the light curve. The
 	double-sine curve was fit to the observed data, excluding the
 	regions of the eclipses ($\chi^{2}_{\nu}$ = 1.6). The dotted
 	lines indicate the width of the primary and secondary
 	eclipses, assuming an accretion disk radius of 0.6$R_{L_{1}}$.
 	} \label{dhump}
\end{figure*}
Figure~\ref{fit} shows the model ellipsoidal variations scaled to the
observed data and Figure~\ref{dhump} is the light curve from which the
model secondary star flux has been subtracted.  The ratio of the mean
secondary flux to the mean observed flux shows that the secondary star
provides approximately 85\% of the observed H-band flux, a value
consistent with the estimate in Szkody \& Mateo
(1986\markcite{szkody1986}) that the disk contributes up to 20\% of
the H-band flux.  The subtracted light curve has both a primary and a
secondary eclipse and shows bright spot ingress and egress features
during primary eclipse.  The peak in the light curve near $\phi$ = 0.8
is early in phase compared to the visible (where it peaks near $\phi$
= 0.9; Wood and Crawford 1986\markcite{wood1986b}).  The gradual
decline from the peak to the start of bright spot ingress is more
extended as a result.  If the peak is due to beamed emission from the
bright spot, its phasing suggests an unusual position for the spot on
the disk.  The subtracted light curve also confirms the presence of a
secondary eclipse of the companion star by the accretion disk.

Unlike the visible data, the near-infrared light curve is not flat
after the primary eclipse, and instead shows a second peak.  It was
this hump which accounted for our failure in fitting the ellipsoidal
variations to this region of the light curve.  The appearance of the
subtracted light curve resembles a double-hump variation such as those
seen in the visible light curves of the dwarf novae WZ Sge and Al Com
(\cite{robinson1978,patterson1996}).  A double-hump variation in IP
Peg would account for the early phase position of the hump before
primary eclipse, and for the presence of the second hump in
Figure~\ref{dhump}.  This phase-dependent variation may be present in
the light curves of other CVs; if so, the double-hump variation may
have been confused with the ellipsoidal variations in other binary
systems.

\subsection{Maximum entropy eclipse mapping of the accretion disk}

The accretion disk was modeled using the maximum entropy eclipse
mapping program developed by Horne (1985) \markcite{horne1985a}.  The
model assumes that a Roche-lobe filling secondary star eclipses a flat
accretion disk which lies in the plane of the binary orbit.  The
maximum entropy method (MEM) divides the surface of the accretion disk
into a two-dimensional grid and numerically solves for the intensity
distribution that best fits the eclipse data while maximizing the
entropy of the model relative to a default map of the disk.  The
program is iterated repeatedly and the default map regularly updated
until the desired quality of fit ($C_{aim} \simeq \chi^{2}_{\nu}$) is
achieved. 

We chose a disk radius equal to $R_{d}$ = 0.56 $R_{L_{1}}$ based on
the variation of the visible bright spot position with time after
outburst (\cite{wolf1993}) and the date of the 1994 September
observations relative to the previous outburst.  After initial
modeling (see below), the disk radius was increased to $R_{d}$ = 0.6
$R_{L_{1}}$.  The initial modeling also demonstrated that fitting the
data to values of $C_{aim} <$ 2.0 introduced spurious features in the
disk map by fitting flickering in the data.  To avoid this, we set the
target fit to $C_{aim}$ = 2.0.  The maximum entropy program assumes
that any variation in the light curve is due to the eclipse,
necessitating the removal of all non-eclipse features from the light
curve before mapping (i.e. the orbital humps and/or the anisotropic
emission from the bright spot).  To remove the double-hump variation,
we fit a double-sine curve to the data, excepting the regions of the
primary and secondary eclipses.  The fit is shown in
Figure~\ref{dhump}.  This figure also shows the extent of the primary
and secondary eclipses in orbital phase for a disk radius of $R_{d}$ =
0.6 $R_{L_{1}}$.  The fit was used to rectify the light curve to a
constant value outside of eclipse.  The rectified light curve is shown
in Figure~\ref{sub}; the flatness of the light curve outside of
eclipse is evidence that the double-sine curve is a good approximation
to the orbital humps in Figure~\ref{dhump}.

Usually, the default maps are smoothed azimuthally so that the maximum
entropy solution favors the most axisymmetric model to fit the
observed data.  This has the effect of suppressing azimuthal
information in the model accretion disk while the radial intensity
distribution remains largely constrained by the data
(\cite{horne1985a}).  The traditional method of choosing an
axisymmetric default map in MEM modeling is ill-suited to eclipses of
IP Peg in quiescence, however, because of the dominant contribution of
the bright spot to the eclipse profile.  We tried many alternative
default maps and modeling schemes to determine the best disk map and
to test the dependence of the map on the method chosen.  Two methods
will be discussed in this text:

1.  We started with a uniform default map of constant intensity and
iterated the program until $C_{aim}$ was reached and the entropy
maximized.  The default map was not updated while iterating.  The
resulting disk map is the least model-dependent map possible, but also
the least physical, subject to 'crossed-arch' distortions of compact
features in the map (\cite{horne1985a}).

2.  We allowed the initially uniform default map to evolve as the disk
map converged on a solution.  The default map was updated regularly by
smoothing the disk map with a Gaussian of width $\sigma$ = 0.01
$R_{L_{1}}$.  The constraint of maximizing entropy then suppresses
disk structure on scales smaller than $\sigma$, while allowing the
data to constrain broad features, producing the smoothest disk map
consistent with the data (\cite{horne1985a}).

\begin{figure*}
	\hspace*{1.60in}\psfig{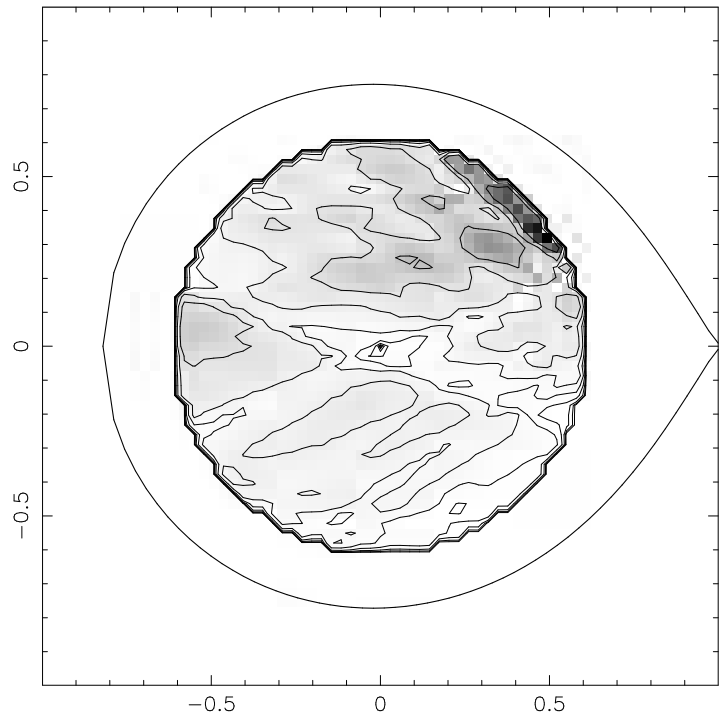}
 	\caption{A contour map of the disk surface brightness, created
using a uniform default map.  The contours are spaced in log intervals
of 0.3.  The x- and y-axes show the dimensions of the accretion disk
in units of $R_{L_{1}}$.}
 	\label{conmap1}
\end{figure*}

\begin{figure*}
	\hspace*{.85in}\psfig{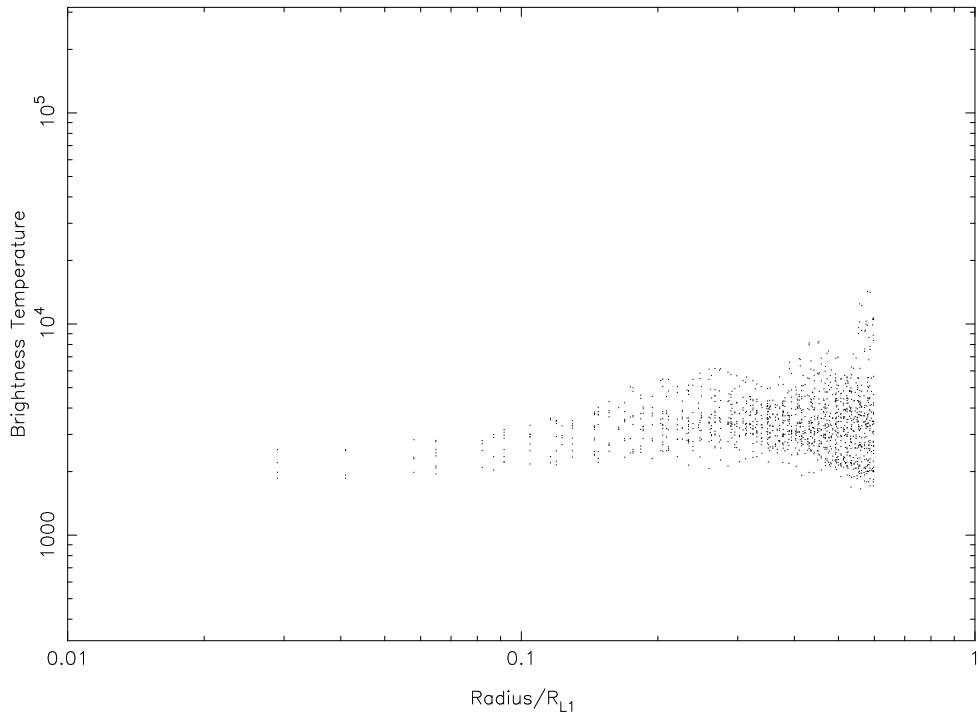}
	\caption{The radial brightness temperature profile of the
	accretion disk in IP Peg, created using a uniform default map.
	The calculation of the disk brightness temperatures assumes a
	distance to IP Peg of 121 pc.}
	\label{brtemp1}
\end{figure*}
\begin{figure*}
	\hspace*{1.60in}\psfig{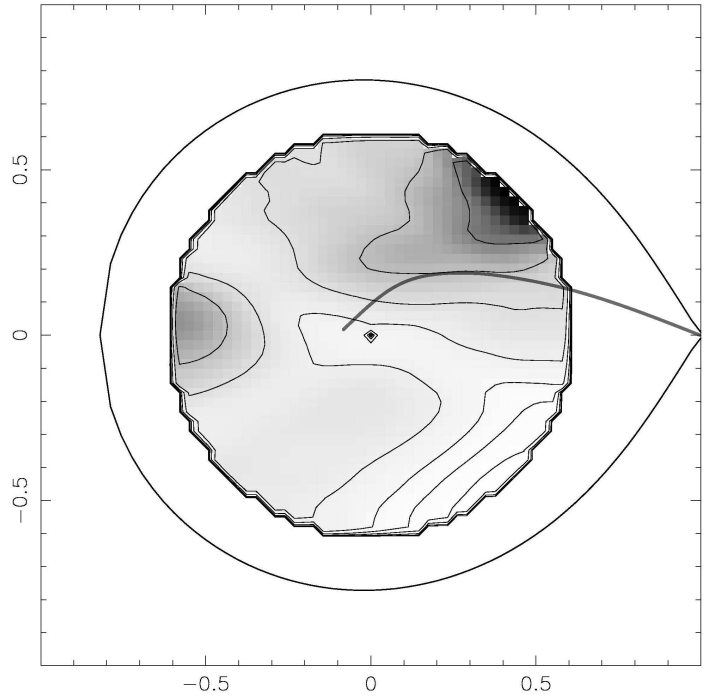}
 	\caption{A contour map of the disk surface brightness, created
using Gaussian smoothed default maps. The contours are spaced in log
intervals of 0.3.  The x- and y-axes show the dimensions of the
accretion disk in units of $R_{L_{1}}$.  Also shown is the theoretical
mass stream trajectory for infalling material from the companion star
(Wood \& Crawford 1986).}
	\label{conmap2}
\end{figure*}

Figure~\ref{conmap1} is the contour map of the accretion disk for the
flat default map.  By assuming blackbody emission at each annulus, we
derived the disk's radial brightness temperature profile, shown in
Figure~\ref{brtemp1}.  We used a distance to IP Peg of 121 pc, based
on the Szkody \& Mateo (1986\markcite{szkody1986}) distance determination but
correcting for the variation in the K-band surface brightness with
(V--K) (\cite{ramseyer1994}).  We set $R_{L_{1}} = 4.5\times10^{10}$
cm (Wood \& Crawford 1986\markcite{wood1986b}, assuming the
intermediate value for the white dwarf boundary layer in Table 4).
Despite the distortions present in the map, several features stand
out.  The accretion disk has a flat radial brightness temperature
profile and a bright spot.  We found the radius of the spot by
creating three disk maps of varying outer disk radius: $R_{d}$ = 0.56,
0.6 and 0.7 $R_{L_{1}}$.  In the latter two cases, the spot intensity
peaked at $x$ = 0.47 and $y$ = 0.34 $R_{L_{1}}$.  This places the
radius of the spot, and thus a lower limit to the disk radius, at 0.58
$R_{L_{1}}$.  This value is larger than the typical radius of the
visible spot during this time in quiescence, so the disk map with an
outer radius $R_{d}$ = 0.56 was too small.  We subsequently set the
disk radius in the maps to $R_{d}$ = 0.6 $R_{L_{1}}$.  The position of
the spot on the disk does not correspond to the range of positions for
the bright spot in IP Peg in visible light; nor does it lie along the
theoretical mass stream trajectory for material being accreted from
the secondary star (the mass stream trajectory is shown in
Figure~\ref{conmap2}; \cite{wood1989b}). 

The disk map in Figure~\ref{conmap1} was also used to determine the
position of $\phi$ = 0 in the observed light curve.  This was done by
shifting the light curve in phase until the intensity contours at
small disk radii and the central 'crossed-arch' distortion were
aligned with the geometric center of the disk.  The corrected time of
$\phi$ = 0 was used to calculate the orbital phasing shown in
Figures~\ref{fsept93} --~\ref{foct94}.

\begin{figure*}
	\hspace*{.85in}\psfig{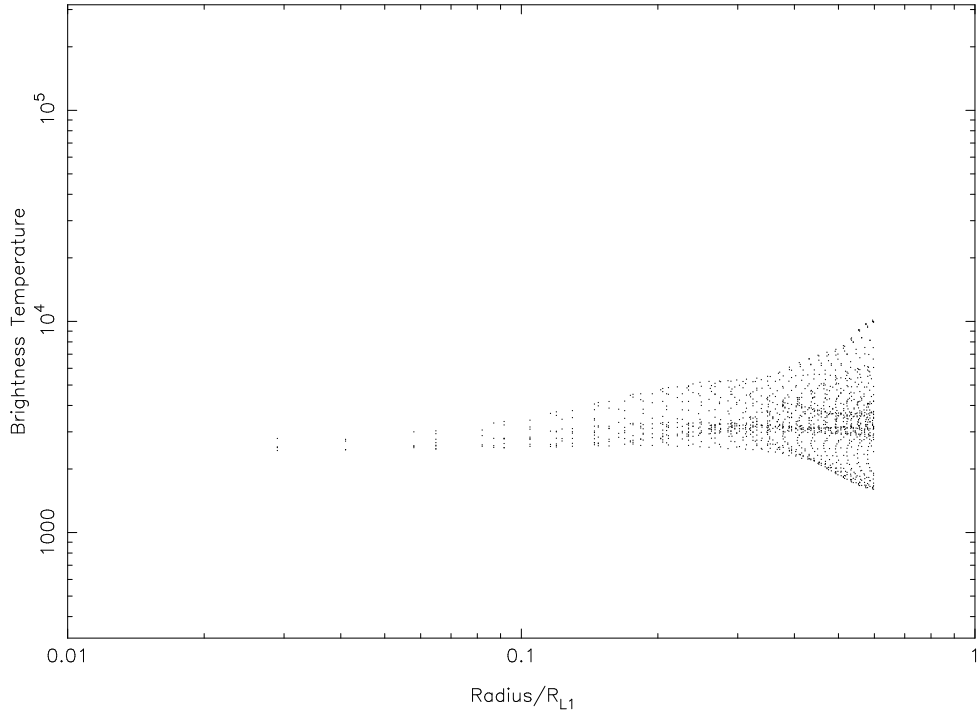}
 	\caption{The radial brightness temperature profile of the
 	accretion disk in IP Peg, created using Gaussian smoothed
 	default maps.  The calculation of the disk brightness
 	temperatures assumes a distance to IP Peg of 121 pc.}
	\label{brtemp2}
\end{figure*}

\begin{figure*}
	\hspace*{.85in}\psfig{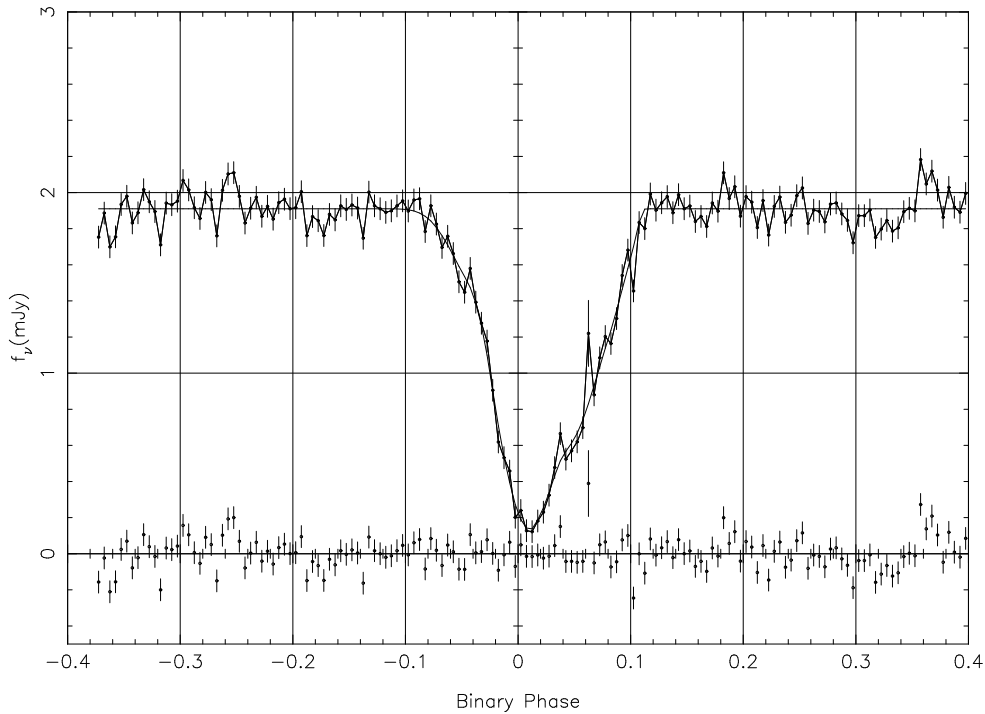}
 	\caption{The primary eclipse of the accretion disk in IP Peg
 	and the fit of the maximum entropy disk map to the data
 	($\chi^{2}_{\nu}$ = 2.0).}
 	\label{modfit}
\end{figure*}

Figure~\ref{conmap2} shows the contour map of the accretion disk
created using the Gaussian-smoothed default map.  Figure~\ref{brtemp2}
is the radial brightness temperature profile for this disk map, and
Figure~\ref{modfit} shows the fit of the disk map to the observed
light curve.   Again, the brightness temperature profile
shows that the bulk of the accretion disk has a flat profile with a
brightness temperature $T_{br} \simeq$ 3000 K.  A bright spot with a
peak temperature $T_{br} \gtrsim$ 10,000 K is located at the edge of
the disk.  The disk map also shows a region of enhanced intensity in
the back of the disk.  In some disk maps, a bright region in the back
of the disk is caused by an underestimate of the unocculted background
flux in the system (\cite{rutten1992b}).  This is not the case in IP
Peg, where alterations in the amount of background light do not affect
this region of the disk map.  Excess light in the back of the disk has
also been indicative of a flare in the opening angles of the disk
(\cite{robinson1995}).  It appears more likely that the enhanced
emission in the map of IP~Peg is an artifact of the maximum entropy
constraint, which favors producing the smoothest disk map possible,
spreading emission from the bright spot over the disk along lines of
constant shadow (\cite{horne1985a}).  The disk map also shows a lower
surface brightness edge at large radii in the fourth quadrant of the
disk in $x-y$ space.  The feature is very robust --- showing up in
every disk map we created --- and is dictated by the shape of the
observed eclipse.
\subsection{Dependence of the results on the modeling parameters chosen.}
Many steps in the modeling of the light curve required fixing
parameters whose true values are uncertain.  Below, we discuss
how the modeling results are affected by changes in several
key parameter values:

1.  The binary geometry, $q$ and $\imath$: The value of the mass ratio
in IP~Peg, $q$, could range from 0.35 to 0.6, although most
observations indicate that $q\geq$ 0.49
(\cite{wood1986b,marsh1988,wolf1993}).  The inclination is most likely
between 79$^{\circ}$ and 81$^{\circ}$.  The disk maps we present
assumed $q$ = 0.49 and $\imath$ = 80$\fdg$9.  We repeated the entire
modeling process for $q$ = 0.6 and $\imath$ = 79$^{\circ}$
(\cite{wolf1993}).  The morphology of the double-hump variation and
the relative depths and shapes of the two eclipses remained the same
in the new light curve, while the peaks in the light curve (near
$\phi$ = 0.25 and 0.75) increased by 0.1 mJy.  The disk maps
consequently showed the same flat brightness temperature distribution
and the prominent bright spot seen in Figures~\ref{conmap1}
and~\ref{conmap2}.  Changing the system geometry did decrease the
brightness temperature in the accretion disk from roughly 3000 K to
2000--2500 K, with a slight rise in temperature with disk radius.  The
peak bright spot temperature declined from 10,000 K to 6000 K,
although much of this decline appears to be due to the flux from the
spot being distributed across the disk (hence the rise in disk
temperature with radius).

2.  The orbital ephemeris and choice of $\phi$ = 0: Given the rapid
changes in IP~Peg's orbital period and the absence of a white dwarf
egress feature in the near-infrared, it is difficult to determine the
correct orbital phasing for the light curve.  We chose the time of
$\phi$ = 0 using the intensity distribution in the accretion disk map.
We tested this assumption by shifting $\phi$ = 0 by 0.01 in phase, so
that the primary eclipse minimum coincides with inferior conjunction
of the secondary star.  Moving $\phi$ = 0 causes small changes in the fits
of the ellipsoidal variations and double-sine curves to the data and
also rotates features in the accretion disk map.  The resulting disk
map had the same intensity distribution, but the intensity contours
were no longer centered on the geometric center of the disk and the
spot was rotated closer to the white dwarf--secondary star axis.  Even
with the rotation, however, the bright spot position did not coincide
with the visible bright spot position or with the theoretical mass
stream trajectory.

3.  Fitting the ellipsoidal model to the data: We scaled the model
ellipsoidal light curve to the data by assuming that 8\% of the
accretion disk flux in Figure~\ref{sub} was unocculted at primary
minimum.  Since the bright spot is a significant source of flux in
IP~Peg, the assumption of uniform disk emission is not strictly true
(see Figure~\ref{conmap2}); however, 0.15 mJy of background light is a
good estimate, and changing the background light by up to 0.05 mJy did
not affect the results.  Larger changes in the background light
(either by eliminating it or by adding large amounts of extra flux)
showed up as artifacts in the disk maps, in particular as light being
removed from or added to the edges of the disk, which
remain unocculted at primary minimum.

4.  Fitting and removing the double-hump variation: Subtracting the
double-sine fit across the primary eclipse (Figure~\ref{dhump})
introduces uncertainty in interpreting the subsequent disk maps.
Without knowing the source of the double-hump variation, and in
particular if it is wholly or partially an anisotropic emitting
source, it is difficult to determine the shape of the rectification
across the primary eclipse.  To test the dependence of the models on
the shape of the fit across the eclipse, we repeated the disk mapping
procedure after using a straight line drawn between the beginning and
end-points of the eclipse to rectify the light curve.  The resulting
accretion disk map was unchanged, with the exception of a slightly
lower peak bright spot temperature (from 10,000 K to 9000 K) and a
larger spot extent on the face of the disk.

5.  The white dwarf: Throughout the modeling, we ignored the
contribution of the white dwarf to the H-band flux.  Marsh
(1988\markcite{marsh1988}) states that the temperature of the white
dwarf is less than 15,000 K, and that it contributes less than 0.2 mJy
in the visible.  We used the light curve synthesis models (discussed
in Section 3.2) to estimate that a 15,000 K white dwarf would
contribute approximately 0.08 mJy to the H-band flux, a small, but not
undetectable, amount (geometric parameters for the white dwarf came
from Wood \& Crawford 1986\markcite{wood1986b}).  The light curves
showed no evidence of white dwarf egress features, however, and the
accretion disk maps did not add light to small disk radii, as would be
expected for unaccounted--for white dwarf flux.  This suggests that
the white dwarf may be much cooler than the upper limit of 15,000 K.

The alternate modeling methods discussed above demonstrate that key
features in the accretion disk map of IP~Peg are robust: the flat
intensity distribution in the disk, the unusual location for the
bright spot, and the brightness temperatures of the accretion disk and
bright spot are relatively unaffected by changes in the modeling
parameters used.

\section{Discussion}
%
%
Figure~\ref{dhump} shows the H-band light curve of IP Peg after the
ellipsoidal variations of the secondary star have been removed.  There
is a double-peaked component to the light curve that mimics but cannot
be attributed to the ellipsoidal variations.  This orbital modulation
is not apparent in the visible light curves of IP Peg, which are flat
following primary eclipse.  The peak before primary eclipse also
occurs at an earlier phase in the NIR than in the visible. While the
visible peak is typically attributed to anisotropic beaming from the
bright spot, the early phase position of the peak in the near-infrared
implies an additional contribution from the double-hump source.
\begin{figure*}
	\hspace*{.85in}\psfig{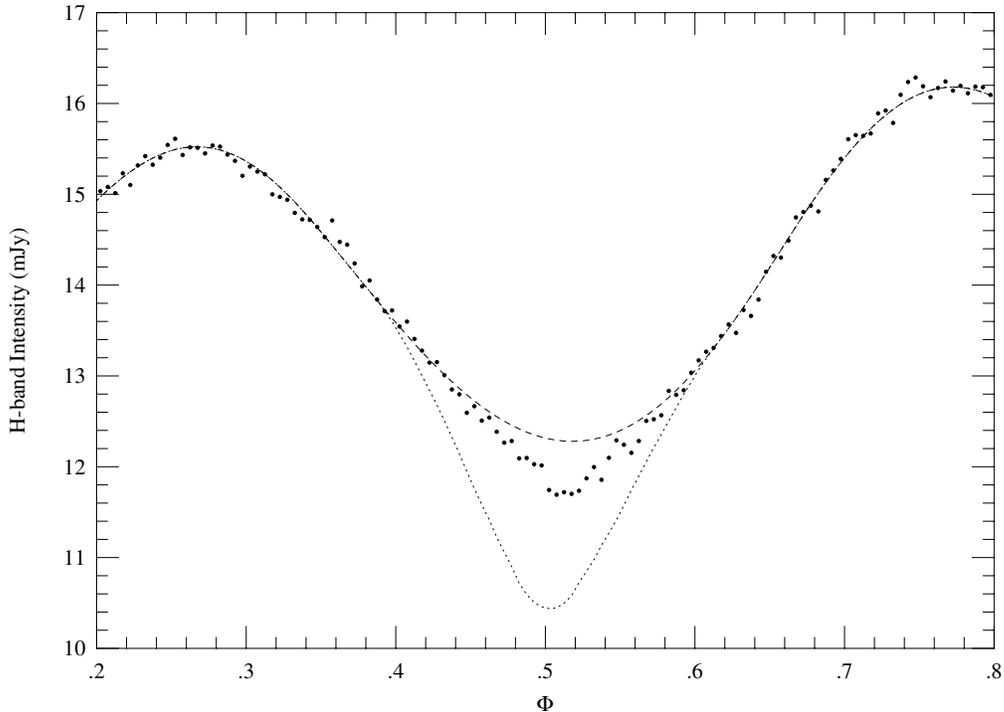}
	\caption{The 1994 September data expanded to show the region
of the secondary eclipse.  The dashed line is the sum of the model
ellipsoidal variations and the double-sine curve used in this paper to
model the secondary star and the double-hump profile.  The dotted line
is the additional amount of occultation expected at secondary eclipse
for a cold, opaque accretion disk ($R_{d}$ = 0.6 $R_{L_{1}}$).  The
observed data fall between these two cases, demonstrating that the
occulting accretion disk is not opaque but is not completely optically
thin, either.  Note that the vertical scaling in this figure has changed
from that in Figure~\ref{fsept94}.}
	\label{allmod}
\end{figure*}
The variation may originate in any of the cool emitters seen in the
near-infrared: the accretion disk, the bright spot, or the secondary
star.  Star spots caused by magnetic activity have been observed on
late-type stars and may preferentially occur at the inner Lagrangian
point in binaries (\cite{ramseyer1995,livio1994}).  A star spot at the
inner Lagrangian point in IP Peg could contribute to the depth of the
light curve at secondary minimum ($\phi$ = 0.5), although a second
persistent spot on the opposite side of the star would also be
necessary to create the observed double-humped variation.

The morphology of the variation observed in IP Peg most strongly
resembles the double-hump profile seen in the quiescent, visible light
curves of WZ Sge and AL Com, two dwarf novae with extremely short
orbital periods.  In these two systems, the double-hump variation
appears to originate in the accretion disk.  In WZ Sge, for example,
the secondary star is virtually invisible even in the near-infrared,
and cannot be the source of the strong variability
(\cite{dhillon1998}).  A double-hump variation is also present in both
systems during the early days of outburst (after which the more common
superhumps are seen), again pointing to a disk origin for the
variability (\cite{patterson1981,patterson1996,kato1996}).  Models of
the double-hump profile in WZ Sge and AL Com have included attributing
it to the bright spot being visible on both the near and far sides of
the accretion disk, or to a spiral dissipation pattern in the disk
caused by a 2:1 resonance instability
(\cite{robinson1978,patterson1996}).  Observations of visible
continuum emission in IP Peg have shown a second hump in the light
curve, which may be caused by the bright spot shining through the disk
(\cite{wolf1998}).  Recent visible spectra of IP Peg on the rise to
outburst and in quiescence have also indicated the presence of
multiple bright emission sites in the accretion disk, which could be
due to spiral structure in the (outbursting) disk or to a second
bright spot from mass-stream overflow (\cite{steeghs1997,wolf1998}).

The presence of the double-hump variation in the NIR light curves of
other CVs cannot be discounted, particularly since the phenomenon may
be inadvertently attributed to ellipsoidal variations from the
secondary star.  It is possible that the contribution of the secondary
star to the light curve could be estimated incorrectly, affecting
determinations of the absolute magnitudes for the secondary star and
the accretion disk.  Ellipsoidal variations have also been used to
constrain the values of $q$ and $\imath$ in other compact binaries.
When these values are used to determine the mass of the primary
object, such as in black-hole binaries, the presence of a variable
disk contribution to the light curve can significantly alter the
calculated results (cf. Sanwal et al.\ 1995, Shahbaz et al.\
1996\markcite{sanwal1995,shahbaz1996}).

The contour map of the quiescent accretion disk in IP Peg
(Figure~\ref{conmap2}) shows that the disk has a flat surface
brightness distribution.  The prominent feature is the bright spot.
The location of the bright spot on the disk is unusual.  It does not
correspond to the range of positions of the bright spot in the visible
nor to the theoretical trajectory of the mass stream from the
secondary star (shown in Figure~\ref{conmap2}).  The near-infrared
bright spot is located at a larger azimuth than the spot seen at
visible wavelengths (where azimuth is measured relative to the line
between the secondary star and the inner Lagrangian point); and it is
at a larger disk radius than is usually seen at this time in
quiescence in the visible, although disk radii this large have been
seen in visible measurements at times soon after outburst
(\cite{wolf1993}).  The disk map also has a lower-intensity edge at
large radii in the part of the disk facing the inner Lagrangian point
in Figure~\ref{conmap2}.  It is unclear why this occurs, but the
feature is robust and unaffected by the method used to create the disk
map or the system parameters chosen.

Except for the spot and turned-down edge, the brightness temperature
of the disk (Figure~\ref{brtemp2}) is flat and clearly deviates from
the $T \propto r^{-3/4}$ law for a steady-state disk.  In this
respect, it resembles the quiescent disk maps of the other eclipsing
dwarf novae, Z Cha, OY Car, and HT Cas, at visible wavelengths
(\cite{wood1986a,wood1989a,wood1992}).  The $T_{br}(r)$ profile of IP
Peg is even flatter than in these systems, however, and the brightness
temperature is lower as well.  The effect could be caused by differing
opacities in different wavebands, or the brightness temperature in the
visible (where the disk in IP Peg has not been mapped) may be lower in
IP Peg than in the other eclipsing dwarf novae.  Temperatures as low
as 2500 K have been seen in OY Car soon after an outburst; the disk
becomes steadily warmer during quiescence (\cite{wood1986c}).  The
observations we modeled were taken a month after the previous outburst
of IP Peg (\cite{aavso287}), so the low temperatures observed are not
likely to be caused by a similar phenomenon.  The flat, non-steady
state brightness temperature profiles seen in these quiescent CVs
provide more support for the disk instability model of normal dwarf
novae outbursts, in which an outburst is caused by a thermal limit
cycle instability in the accretion disk (e.g. Cannizzo
1993\markcite{cannizzo1993}).

The presence of the secondary eclipse in the near-infrared light
curves limits the interpretation of the models of the primary eclipse.
The secondary eclipse indicates that the accretion disk is occulting
some of the flux from the secondary star, so the observed
near-infrared disk flux cannot completely originate in an optically
thin disk.  The secondary eclipse is too shallow to be caused by a
fully opaque disk, however.  We tested the expected depth of the
secondary eclipse for an optically thick accretion disk by allowing a
dark, opaque disk to occult the secondary star (using the light curve
modeling code discussed in Section 3.2).  For an accretion disk radius
of $R_{d}$ = 0.6 $R_{L_{1}}$, the secondary hump at $\phi$ = 0.5 would
be 1.85 mJy deeper than it is without the secondary eclipse; the
effect of adding the secondary eclipse is shown in
Figure~\ref{allmod}.  This depth is much larger than the dip at $\phi$
= 0.5 in Figure~\ref{dhump}, even before the double-sine fit is
subtracted.  A smaller disk radius of $R_{d}$ = 0.56 $R_{L_{1}}$ (the
original guess based on visible observations) occults 1.7 mJy at
secondary eclipse, still too deep to be consistent with the observed
hump at $\phi$ = 0.5.  The shallow secondary eclipse also rules out a
two-phase accretion disk (an opaque, cool disk with a hot, optically
thin chromosphere), unless the size of the opaque disk is smaller than
the disk radius inferred from the observed visible and near-infrared
bright spot positions.

The shallow secondary eclipse shows that optically thin emission
dominates the observed H-band flux from most of the disk in IP~Peg.
This result is consistent with the near-infrared colors of IP~Peg,
which, when plotted on a flux-ratio diagram point to a strong
optically thin component to the flux (\cite{szkody1986,berriman1985}).
The distribution of the transparent and opaque gas is unclear,
however.  The disk emissivity could be patchy.  In particular, while
the accretion disk is primarily optically thin, the bright spot
emission is probably optically thick because its brightness
temperature is greater than 10,000 K, and is near the temperature of
the spot determined from visible observations (\cite{marsh1988}).

The effective temperature of the accretion disk is a more complex
issue.  For a one-phase disk, either completely opaque or transparent,
the radial brightness temperatures in Figure~\ref{brtemp2} give a
rough lower limit to the effective temperature of the disk.  For a
transparent disk, the brightness temperature is likely to be a severe
underestimate of the actual disk temperature.  Since the secondary
eclipse in IP~Peg is too shallow to allow for an opaque accretion
disk, models for the quiescent disk which have temperatures of
$T_{eff} \simeq$ 5000 -- 6000 K (and are optically thin) are more
consistent with the data than models which invoke a cold ($T_{eff}
\simeq$ 2000 -- 3000 K), optically thick disk.

The 1993 September and 1994 October light curves were too noisy to
model.  It is clear, however, particularly from the variations in the
shape of the light curves in the month from 1994 September to October,
that parameters in the accretion disk in IP~Peg---the amplitude of the
double-hump variation, and the temperatures and densities in the disk
and the bright spot---vary over the course of the quiescent cycle.  As
a result, additional and repeated modeling of the double-hump
variation and both eclipses would better resolve the optical depth and
temperature of the quiescent disk in IP Peg.  Future work should also
include simultaneous, multicolor eclipse maps in the near-infrared to
further constrain the optical depth and temperature of the disk by
probing how the NIR colors vary throughout the binary orbit.

\section{Conclusions}
%
%

1.  The quiescent H-band light curve of IP Peg contains contributions
from the late-type companion star, the accretion disk and the bright
spot, as well as a primary eclipse of the accretion disk and a
secondary eclipse of the companion star.  The characteristic
ellipsoidal variations from the secondary star dominate the light
curve, but the amplitude of the variation is not enough to account for
all of the orbital modulation seen in the light curve outside of
eclipse.

2.  The light curve after the model secondary star flux has been
subtracted shows a phase-dependent, double-hump profile reminiscent of
the quiescent light curves of WZ Sge and AL Com.  In these two
systems, the double-hump variation is believed to originate in the
accretion disk.  The presence of a double-hump variation in the
near-infrared light curves of other compact binaries may complicate
determinations of the relative flux contributions of the accretion
disk and the secondary star.

3.  The primary eclipse was modeled using maximum entropy eclipse
mapping techniques.  The bulk of the disk has a flat surface
brightness distribution and a cool brightness temperature ($T_{br}
\simeq$ 3000 K).  There is a prominent bright spot on the edge of the
disk ($T_{br} \simeq$ 10,000 K).  The position of the near-infrared
bright spot is not the same as the position of the theoretical mass
stream trajectory or the range of measured visible bright spot
positions in IP~Peg.

4.  The flat radial brightness temperature distribution of the
accretion disk is consistent with those of other eclipsing dwarf novae
in quiescence, although the near-infrared disk in IP Peg is both
flatter and cooler than in the other systems.  The flat intensity
distribution is consistent with the quiescent, non-steady-state
behavior predicted by the disk instability model of normal dwarf novae
outbursts.

5. The secondary eclipse of the companion star indicates that some
occulting material is present in the disk, but the eclipse depth is
too shallow to allow for an opaque accretion disk.  The disk
emissivity could be patchy.  In particular, while optically thin
emission dominates the H-band flux, the bright spot is probably
optically thick with a temperature around 10,000 K.  The temperature
of the bulk of the accretion disk depends on its optical depth, but
is probably higher than the brightness temperature.

\acknowledgments{The authors wish to thank Chris Johns-Krull for
generously providing the Allard model atmospheres for M dwarf stars
and the software to generate model atmospheres of IP Peg.}

\end{document}